\documentclass[11pt]{article}
\pdfoutput=1
\usepackage{graphicx}
\usepackage{hyperref}
\usepackage{epsfig}
\usepackage{amsmath}
\usepackage{amsthm}
\usepackage{amssymb}
\usepackage{multirow}
\usepackage{color}
\usepackage[nosort]{cite}
\usepackage{hyperref}
\usepackage[fontsize=11.45pt]{scrextend}
\usepackage{braket,mathrsfs,slashed}
\usepackage{float}
\usepackage{physics}
\usepackage{setspace}

\usepackage[caption=false]{subfig}
\newlength{\abstractwidth}
\newcommand{\be}{\begin{equation}}
\newcommand{\ee}{\end{equation}}
\renewcommand{\title}[1]{\vbox{\center\bf{\Large{#1}}}\vspace{5mm}}
\renewcommand{\author}[1]{\vbox{\center#1}\vspace{5mm}}
\newcommand{\address}[1]{\vbox{\center\em#1}}

\usepackage{dsfont}

\usepackage[utf8]{inputenc}
\usepackage{rotating}
\usepackage{dsfont}
\usepackage[paper=a4paper,margin=1.5cm]{geometry}
\usepackage[utf8]{inputenc}
\usepackage{rotating}
\usepackage{soul}

\usepackage{mathrsfs} 
\usepackage{bbm}
\usepackage{etoolbox} 
\usepackage{multirow}

\renewcommand\[{\begin{equation}}
\renewcommand\]{\end{equation}}

\newcommand{\ba}{\begin{eqnarray}}
\newcommand{\ea}{\end{eqnarray}}

\hypersetup{pdfstartview=FitV,colorlinks=true,linkcolor=midblue,citecolor=midblue,filecolor=midblue,urlcolor=midblue}

\definecolor{midblue}{rgb}{0,0,0.5}

\begin{document}
	
		\newgeometry{top=3.1cm,bottom=3.1cm,right=2.4cm,left=2.4cm}
		
	\begin{titlepage}
	\begin{center}
		\hfill \\
		\vskip 0.5cm

		\title{Weak Gravity Conjecture in the sky: \\
        
        \vspace{4pt}
        gravitational waves from preheating in \\
        
        \vspace{7pt}
        Einstein-Maxwell-Scalar EFT}

			\author{\large Jiaxin Cheng$^{a,b,c\,\ddagger}$, Anna Tokareva$^{a,d,e\,\dagger}$ }
			
			\address{$^a$School of Fundamental Physics and Mathematical Sciences, \\Hangzhou Institute for Advanced Study, UCAS, Hangzhou 310024, China\\[1.5mm]
                
                $^b$Institute of Theoretical Physics, Chinese Academy of Sciences, Beijing 100190, China\\[1.5mm]
                $^c$University of Chinese Academy of Sciences, Beijing 100049, China\\[1.5mm]
                $^d$Department of Physics, Blackett Laboratory, Imperial College London, SW7 2AZ London, UK\\
                $^e$International Centre for Theoretical Physics Asia-Pacific, Beijing/Hangzhou, China\\[1.5mm]}
				\vspace{.3cm}

		\end{center}

\vspace{0.15cm}

\begin{abstract}
The effective field theory (EFT) concept provides a necessary tool for obtaining general predictions of low-energy theory valid below its unitarity-breaking scale (cutoff scale). Early Universe inflation and subsequent reheating could be a unique setup for testing potentially observable effects coming from the derivative expansion of the corresponding EFT around the flat space vacuum. In this work, we consider an EFT describing perturbative reheating dominated by the decay of inflaton to photons caused by the dimension-5 operator $\phi F_{\mu\nu} F^{\mu\nu}$. We compute the graviton production during reheating and high frequency gravitational wave signal due to the bremsstrahlung effect in the presence of $R_{\mu\nu\lambda\rho}F^{\mu\nu} F^{\lambda\rho}$ operator. It may lead to the dominant contribution at high momenta if the EFT cutoff is lower than the Planck mass. Assuming the general consequences of the unitarity and causality constraints, which imply that all EFT operators should be present, and be suppressed by the scales following from the dimension analysis, we obtain the observational constraints (CMB bound for the dark radiation) on the mass of the inflaton and UV cutoff of gravity. We find that for the typical parameters of large field inflation models, the gravitational cutoff scale cannot be lower than $10^{15}$ GeV.
\end{abstract}
\vspace{4cm}
\noindent\rule{6.5cm}{0.4pt}\\
$\,^\dagger$ \href{mailto:tokareva@ucas.ac.cn}{tokareva@ucas.ac.cn}\\
$\,^\ddagger$ \href{mailto:chengjiaxin24@mails.ucas.ac.cn}{chengjiaxin24@mails.ucas.ac.cn}

\end{titlepage}

{
	\hypersetup{linkcolor=black}
	\tableofcontents
}

\baselineskip=17.63pt



\newpage

\section{Introduction and summary}
\label{sec:intro}
The inflationary scenarios originally proposed in \cite{STAROBINSKY198099,GuthPhysRevD.23.347,LINDE1982389,Mukhanov:1981xt} postulate that the universe underwent an exponential expansion before the hot Big Bang epoch. This assumption provides a natural explanation of the homogeneity of the present-day Universe, as well as a parameterization of cosmological perturbations observed in CMB \cite{Planck:2018jri}. 
Comprehensive reviews of different inflationary models can be found in \cite{Olive:1989nu,Lyth:1998xn,Martin:2013tda,Ellis:2023wic,kallosh2025presentstatusinflationarycosmology}. At the current stage, the choice of the particular model cannot be made just based on the observational data, although the constraints on inflationary parameters are getting stronger \cite{AtacamaCosmologyTelescope:2025blo,SPT-3G:2025bzu,DESI:2025zgx}. 

 The inflationary stage can be dynamically realized in a microscopic model containing a scalar field with a potential allowing for the slow-roll regime. In this regime, inflation is driven by a scalar field $\phi$, whose action can be written as 
\begin{align}
    S=\int d^4{x}\sqrt{-g}\left(-\frac{1}{2}\partial_{\mu}\phi\partial^{\mu}\phi-V(\phi)\right).
\end{align}
Inflation ends by the stage of fast rolling down of the field towards the minimum of its potential and subsequent oscillations around it. These oscillations generate the Standard Model particles. This process, usually called reheating, can be described as the decay or scattering of the inflatons to the other particles. The reheating stage leads to intense production of high-frequency gravitons. If it admits a description in a perturbation theory, the main sources of gravitons are inflaton decay, bremsstrahlung and scattering of inflatons to gravitons \cite{Barman:2023ymn,Barman:2023rpg,Bernal:2023wus,Bernal:2025lxp,Garcia:2020wiy,Ringwald:2022xif,Mondal:2025kur,Choi:2024ilx,Ema:2021fdz,Garcia:2023eol,Klose:2022knn,Jiang:2024akb,Das:2025cqs,Xu:2024fjl,Barman:2024htg}. These gravitons would freely propagate in the Universe as extra dark radiation component. For this reason, models predicting too efficient production of gravitons during reheating can be ruled out by the CMB bound on the number of extra relativistic degrees of freedom \cite{Planck:2018vyg}.

Inflation and reheating are the phenomena happening at high energy, where the underlying theoretical description of both matter and gravity is not well established. Even if one makes an extra assumption that both the Standard Model (SM) and General Relativity (GR) work up to the Planck scale, GR is a non-renormalisable theory. For this reason, the overall setup can be described only as an Effective Field Theory (EFT) \cite{Burgess:2017ytm}. In order to describe gravion production during reheating, we construct all relevant EFT operators that can contribute to the process of inflaton decay with emission of an extra graviton (bremsstrahlung). The action describing gravitons can be written in the form of an expansion in powers of the Weyl tensor and its covariant derivatives. This expansion starts from
\begin{equation}
\label{GREFT}
    S_{GR} = M_{P}^{2} \int d^4 x \sqrt{-g} \left\{
-\frac{R}{2} +  \frac{a_3}{\Lambda_{\text{UV}}^4} [{\cal C}^3] + \frac{a_4}{\Lambda_{\text{UV}}^6} C^2  + \frac{\tilde{a}_4}{\Lambda_{\text{UV}}^6} \tilde{C}^2 + \cdots
\right\},
\end{equation}
where $[{\cal C}^3]=C_{\mu\nu\lambda\rho}C^{\lambda\rho\alpha\beta} C_{\alpha\beta}^{\mu\nu}$, $C = C_{\mu\nu\rho\sigma} C^{\mu\nu\rho\sigma}$ and $
\tilde{C} = C_{\mu\nu\rho\sigma} \tilde{C}^{\mu\nu\rho\sigma}$, and $\tilde{C}_{\mu\nu\rho\sigma} = \frac{1}{2} \epsilon_{\mu\nu\alpha\beta} C_{\alpha\beta\rho\sigma}$. A number of terms in this expansion were not written because, by means of perturbative field redefinitions, they can be reduced to \eqref{GREFT} \cite{Ruhdorfer:2019qmk,deRham:2022gfe,Basile:2024oms}. If one considers an EFT describing gravitons and photons (Einstein-Maxwell EFT), the first terms in the derivative expansion can be written as \cite{Davies:2021frz,Henriksson:2022oeu}
\begin{equation}
\label{EM EFT}
    S_{EM} = \int d^4 x \sqrt{-g} \left[
-\frac{M_P^2}{2}\,R - \frac{1}{4} F_{\mu\nu} F^{\mu\nu} + \frac{\beta}{\Lambda_2^2} R_{\mu\nu\rho\sigma} F^{\mu\nu} \tilde{F}^{\rho\sigma} + \frac{\gamma}{\Lambda_{UV}^4} (F^{\mu\nu} F_{\mu\nu})^2 +  \frac{\tilde{\gamma}}{\Lambda_{UV}^4} (F_{\mu\nu} \tilde{F}^{\mu\nu})^2+\cdots
\right].
\end{equation}
In this work, we study reheating caused by the inflaton decay to gauge fields, focusing on the example of photons. We assume the main decay channel for the inflaton is 
\begin{equation}
    L_{\phi\gamma\gamma}=\frac{\alpha}{\Lambda_{1}}\phi F_{\mu\nu}F^{\mu\nu},
\end{equation}
This process is followed by graviton production dominated by bremsstrahlung. Thus, in the EFT framework, the first terms in the action describing interactions of the scalar field, photon, and gravity are
\begin{equation}
    L_{\phi}=\frac{1}{2}(\partial_{\mu}\phi)^2-\frac{1}{2}m^2 \phi^2+\frac{\alpha}{\Lambda_{1}}\phi F_{\mu\nu}F^{\mu\nu}+\frac{\delta}{\Lambda_3}\phi\,R_{\mu\nu\rho\sigma}R^{\mu\nu\rho\sigma}+\cdots.
\end{equation}
The last term causes the effect of inflaton decay to two gravitons, which was studied in \cite{Ema:2021fdz,Ema:2020ggo,Tokareva:2023mrt,Koshelev:2022wqj}. In this work, we mainly focus on computing the three-particle decay of the inflaton to two photons and a graviton, leading to the high-frequency graviton production. In this way, we probe the coupling $\beta$ relating it to the potentially observable quantities, such as the gravitational wave signal. Although the direct detection of high-frequency gravitational waves is challenging \cite{Aggarwal:2020olq,Aggarwal:2025noe,Guo:2025cza}, the CMB bound on their contribution to the number of extra relativistic degrees of freedom \cite{Planck:2018vyg,Yeh:2022heq} is capable of putting severe constraints on EFT couplings and unitarity breaking scale. Given the expected one order of magnitude improvement of the existing constraints in future experiments, such as LiteBIRD \cite{LiteBIRD:2022cnt}, Simons Observatory \cite{SimonsObservatory:2018koc,SimonsObservatory:2019qwx}, and CMB-S4 \cite{CMB-S4:2016ple}, the dark radiation constraint can become a window to the physics of reheating, at the same time probing the description of the Universe intriguingly close to the Planck scale.

Although it is usually assumed that the theory coupled to gravity breaks down at the Planck scale, it may also be the case that this scale is lower. For example, the presence of a large number of light states can significantly decrease the unitarity breaking scale for gravitational EFT \cite{0706.2050,0710.4344,0806.3801,0812.1940,Caron-Huot:2024lbf} (species bound). If the EFT can be trusted only up to the scale $\Lambda_{UV}$, the other higher derivative operators are suppressed by certain combinations of $\Lambda_{UV}$ and Planck mass. The precise scaling of each Wilson coefficient in a viable EFT in principle can be determined from causality and unitarity constraints (EFT bootstrap) \cite{Bellazzini:2019xts,Arkani-Hamed:2021ajd,Henriksson:2021ymi,Alberte:2020bdz,Saraswat:2016eaz,Guerrieri:2021ivu,Pham:1985cr, Pennington:1994kc,Nicolis:2009qm, Komargodski:2011vj,Remmen:2019cyz,Herrero-Valea:2019hde,Bellazzini:2017fep,deRham:2017avq,deRham:2017zjm,deRham:2017imi,Wang:2020jxr, Tokuda:2020mlf,Li:2021lpe,Caron-Huot:2021rmr,Du:2021byy,Bern:2021ppb,Li:2022rag, Caron-Huot:2022ugt,Herrero-Valea:2020wxz,EliasMiro:2022xaa,Bellazzini:2021oaj, Sinha:2020win, Trott:2020ebl,Herrero-Valea:2022lfd,Hong:2023zgm,Chiang:2022jep,Huang:2020nqy,Noumi:2021uuv, Xu:2023lpq, Chen:2023bhu,Noumi:2022wwf,deRham:2022hpx, Hong:2024fbl,Bern:2022yes,Ma:2023vgc,DeAngelis:2023bmd,Acanfora:2023axz,Aoki:2023khq,Xu:2024iao,EliasMiro:2023fqi,McPeak:2023wmq,Riembau:2022yse,Caron-Huot:2024tsk,Caron-Huot:2024lbf,Wan:2024eto,Buoninfante:2024ibt,Berman:2024owc,deRham:2025vaq,Berman:2025owb,Nie:2024pby,Bellazzini:2021oaj,Riembau:2022yse,Bellazzini:2020cot,Beadle:2024hqg,Ye:2024rzr,Bertucci:2024qzt,Chang:2025cxc, Beadle:2025cdx, Pasiecznik:2025eqc,Ye:2025zhs}, although in realistic models the full consideration has not been done. In general, the obtained bounds coincide with naive expectations from the power counting \cite{Hong:2023zgm}. For the EFT couplings relevant for our work, it implies the natural expectations
\begin{equation}
    \Lambda_1=\Lambda_{UV}, \quad \Lambda_2=\frac{\Lambda_{UV}^3}{M_P^2} , \quad |\alpha| \sim 1, \quad |\beta|\sim 1.
\end{equation}
The scaling of $\Lambda_2$ was also more rigorously obtained in \cite{Henriksson:2022oeu}. It is interesting to mention that the consistency constraints obtained from the Weak Gravity Conjecture (WGC) for black hole evaporation \cite{Bellazzini:2019xts,Arkani-Hamed:2021ajd,Arkani-Hamed:2006emk,Cheung:2018cwt,Cottrell:2016bty,Hebecker:2017uix,Abe:2023anf,DeLuca:2022tkm,Barbosa:2025uau,Cao:2022ajt,Cao:2022iqh,Aalsma:2019ryi} imply higher values of $\Lambda_2 \sim \Lambda_{UV}^2/M_P$, thus leading to less efficient graviton production. In this way, such a physical observable as the high-frequency gravitational waves spectrum appears to be sensitive to the refined EFT structure determined by the consistency conditions and, additionally, string Swampland conjectures \cite{Vafa:2005ui}.

We compute the differential decay rate for the inflaton producing the gauge fields with an emission of a single graviton. We focus on the case of reheating due to the inflaton decay to photons, but our results can be straightforwardly generalised to the case of $W,~Z$-bosons and gluons. The goal of our work is to parametrize the resulting GW signal by the reheating temperature, EFT cutoff, and inflaton mass. Requiring that the gravitational wave signal doesn't exceed the CMB bound \cite{Planck:2018vyg}, we obtained a constraint on the EFT cutoff scale. We found that the CMB bound on dark radiation constrains the UV cutoff scale of gravitational EFT describing reheating to be higher than $10^{16}$ GeV for values of inflaton mass larger than $10^{12}$ GeV, typical for large field inflation models. 

The paper is organised as follows.
In Section~\ref{sec:intro}, we outline the EFT framework for describing inflaton decay and graviton bremsstrahlung.
Section~\ref{sec:eft_setup} presents the setup of the EFT for reheating, including the relevant interactions between the inflaton, photons, and gravitons.
In Section~\ref{sec:reh_unitarity}, we discuss the general expected structure of consistent EFTs and motivate the choice of suppression scales for different operators.
Section~\ref{sec:graviton_prod} is devoted to the computation of graviton production during reheating, where we derive the differential decay rate, solve the Boltzmann equation, and obtain the gravitational wave spectrum.
The observational constraints on the UV cutoff scale, derived from the CMB bound on dark radiation, are presented in Section~\ref{sec:constraints}, for the cases when one insists on the WGC constraint and when violations of this condition are still allowed.
Finally, we summarise our findings and discuss their implications in Section~\ref{sec:conclusions}.

\section{Set-Up of the EFT of reheating}
\label{sec:eft_setup}
Let us list here the EFT couplings between the inflaton, the graviton, and the photon, which are responsible for reheating and graviton bremsstrahlung.

	\begin{align}
    \label{action}
    	S_{int}=\int d^4 x\sqrt{-g}\left(\mathcal{L}_{\phi\phi }+\mathcal{L}_{\phi\gamma\gamma}+\mathcal{L}_{\gamma\gamma h}\right),
	\end{align}
	\begin{align}
    	\mathcal{L}_{\phi\phi }&=\frac{1}{2}\partial_{\mu}\phi\partial^{\mu}\phi-\frac{1}{2}m^2\phi^2,\\
    	\mathcal{L}_{\phi\gamma\gamma}&=\frac{\alpha}{\Lambda_1}\phi F_{\mu\nu}F^{\mu\nu},\\
    	\mathcal{L}_{\gamma\gamma h}&=-\frac{1}{4}F_{\mu\nu}F^{\mu\nu}+\frac{\beta}{\Lambda^2_2}R_{\mu\nu\rho\sigma}F^{\mu\nu}F^{\rho\sigma}.
	\end{align}
We make the following assumptions about the cosmological scenario of the early Universe: 
\begin{enumerate}
		\item Inflationary stage is driven by the scalar field $\phi$ with a potential $V(\phi)\approx m^2\phi^2/2$ for small field values relevant for reheating. Our results don't depend on a particular choice of the inflation model, although they are mainly relevant for the large field inflation with the Hubble scale of order $H_{inf}\sim 10^{11}\, - \, 10^{14}$ GeV. This setting also implies the bound on maximal reheating temperature, $T_{reh}<6.6\cdot 10^{15}$ GeV \cite{Ringwald:2020ist}.
		\item At the end of inflation, the scalar field $\phi$ oscillates with dissipation at the bottom of its potential $V(\phi)\approxeq \frac{1}{2}m^2\phi^2$, finally transferring the energy of the background field to the thermal bath of Standard Model particles.
        \item Reheating can be described as a perturbative decay of the inflaton field.
		\item At the reheating stage, the channel $\phi\to 2\gamma$ is the dominant channel responsible for the energy transfer to the SM particles before thermalization.
		\item We assume that the reheating is described by the EFT of inflation and gravity, as well as their interaction with matter.
        \item We assume that all the relevant interactions obey parity conservation.
	\end{enumerate}

The computations performed in this paper are done for the case when the inflaton decays to photons is responsible for reheating. However, this setup can be straightforwardly generalised to the case when the inflaton decays to the SM gauge fields or gluons. In this sense, the results and constraints obtained in this work can be stated as the ones expected in more generic models, given that the reheating stage happens due to the perturbative decay of the inflaton to the SM particles. As the coupling of all particles to gravity is universal (up to order one multiplies), we expect the gravitational wave signals to have the same order of magnitude. If reheating is happening due to non-perturbative effects, such as parametric resonance \cite{del-Corral:2024vcm,Palma:2000md,Henriques:2001zt,Sfakianakis:2018lzf,Kaiser:1997hg,Greene:1997fu}, tachyonic instability \cite{Kofman:2001rb,Bezrukov:2019ylq,Hashiba:2021gmn,He:2020ivk,Klose:2022rxh,Rubio:2019ypq}, or includes large inhomogeneities in the inflaton condensate \cite{hep-ph/9701423,astro-ph/0601617,gr-qc/9909001,1707.04533,2206.14721,2312.15056,Piani:2023aof}, the production of gravitons should be even more intense. Such models were widely studied in the literature with the use of numerical simulations under the assumptions that $\Lambda_{UV}= M_P$ and, thus, the effects of higher derivative couplings can be neglected \cite{2312.15056}.

\section{Reheating and unitarity breaking scale}
\label{sec:reh_unitarity}
\subsection{On the choice of the scale $\Lambda_1$ and reheating temperature}

\begin{figure}[htbp]
    \centering
    \includegraphics[width=0.3\linewidth]{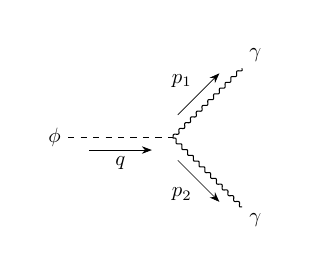}
    \caption{Feynman diagram of inflaton decays to two photons at the leading order. Here $\phi$ is the inflaton, and $\gamma$ stands for a photon. The variables $q$ and $p_1, p_2$ are the momenta of the inflaton and photons, respectively.}
    \label{fig:decay rate}
\end{figure}
According to our assumptions, reheating is dominated by the channel $\phi\to 2\gamma$ which is described by the Lagrangian $\mathcal{L}_{\phi\gamma\gamma}$. Given that the decay is happening in a perturbative regime, we can calculate the process $\phi\to \gamma\gamma$ at the leading order. The corresponding Feynman diagram is shown in Figure \ref{fig:decay rate}, whose matrix elements are 
\begin{align}
    &i\mathcal{M}_{++}=i\mathcal{M}_{--}=-4i\frac{\alpha m k}{\Lambda_1},\quad i\mathcal{M}_{+-}=i\mathcal{M}_{-+}=0.
\end{align}
The total decay rate can be obtained as
\begin{align}
    \Gamma_{\phi\gamma\gamma}=\frac{m^3\alpha^2}{4\pi\Lambda_1^2}.
\end{align}
If we assume that reheating is dominated by the decay of inflaton to photons, we can determine the moment of reheating as $H=\Gamma_{\phi\gamma\gamma}$. Thus, we have 
\begin{align}
    \left\{
    \begin{aligned}
        &\rho=\frac{\pi^2}{30}g_{\ast}T^4,\\
        &H_{reh}^2=\frac{1}{3M_{p}^2}\rho,
    \end{aligned}\right.
\end{align}
 Here $g_{\ast}$ is an effective number of relativistic degrees of freedom at the moment of reheating. In the Standard Model, we have $g_{\ast}=106.75$. The first equation comes from the fact that right after the reheating stage, the Universe is radiation dominated, and the second equation is the Friedmann equation. Thus, one can obtain the reheating temperature as 
\begin{align}
    T_{reh}=\left[\frac{90}{\pi^2 g_{\ast}}\right]^{\frac{1}{4}}\sqrt{M_{P}\Gamma_{\phi\gamma\gamma}}=\left[\frac{90}{\pi^2 g_{\ast}}\right]^{\frac{1}{4}}\sqrt{\frac{m^3\alpha^2 M_P}{4\pi\Lambda_1^2}}.
\end{align}
The reheating temperature is fully determined by the inflaton mass and the scale $\Lambda_1$ (if we set $\alpha=1$), so we can revert this expression and parametrize the model only by the reheating temperature and inflaton mass,
\begin{equation}
    \frac{\Lambda_1}{\alpha}=\left(\frac{90}{\pi^2 g_{\ast}}\right)^{\frac{1}{4}}\sqrt{\frac{M_P m^3}{4\pi T_{reh}^2}}.
\end{equation}
Furthermore, if we assume that the unitarity breaking scale of the theory is $\Lambda_{UV}$, the scale $\Lambda_1$ should be larger, $\Lambda_1\gtrsim\Lambda_{UV}$, otherwise this operator would cause the EFT breakdown at the scale below $\Lambda_{UV}$. In principle, it is still possible that $\Lambda_1\gg \Lambda_{UV}$, i.e., this coupling is additionally suppressed compared to the one expected from the power-counting. However, it is not a natural choice of parameters, as this coupling is expected to be generated with $O(1)/\Lambda_{UV}$ through loop corrections from the other operators. It is hard to preserve this kind of fine-tuning unless there is an approximate symmetry protecting such a coupling. In addition, arguments based on unitarity and causality \cite{Hong:2023zgm} also support the EFT structure with operators suppressed by a single $\Lambda_{UV}$ scale. For these reasons, throughout this paper we focus on the case when $\Lambda_1=\Lambda_{UV}$ which corresponds to the maximal reheating temperature, given the value of $m$ is chosen. This choice would also correspond to the most efficient graviton production, thus allowing us to explore the maximal possible gravitational wave signal at high frequencies. In further considerations, we will also take into account such obvious consistency requirements as 
\begin{equation}
    m\ll\Lambda_{UV},\qquad T_{reh}\ll\Lambda_{UV},
\end{equation}
which guarantees that the reheating stage can be described at all by the EFT.

\subsection{Scale $\Lambda_2$: unitarity bound and Weak Gravity Conjecture}

The scale $\Lambda_2$ in general may not parametrically coincide with the scale $\Lambda_{UV}$\footnote{They are of the same order, $\Lambda_2\sim\Lambda_{UV}$, for example, if the EFT couplings emerge from integration out either a single or a few heavy states with $m_{UV}\sim\Lambda_{UV}$, while after adding them back the theory becomes renormalizable until Planck scale. The similar scaling emerges, for example, in the Euler-Heisenberg Lagrangian.}. If we assume that $\Lambda_2$ is such that the coupling $R F F$ saturates the unitarity bound, then
\begin{equation}
\label{unitarity}
\Lambda_2\simeq\frac{\Lambda_{UV}^3}{M_P^2}.
\end{equation}
This scaling can be obtained from a naive power counting, considering the unitarity breaking scale for $\gamma\gamma\rightarrow\gamma\gamma$ amplitude, where the photon-graviton coupling is dominated by $\Lambda_2$ operator. In a more rigorous way, such parametric dependence (in the case that the constraint is saturated) can also be extracted from the results of the EFT bootstrap for photons and gravitons, see Figures 5 and 11 of \cite{Henriksson:2022oeu}. This conclusion can be made if the suppression scale of $F^4$ operators is $\Lambda_{UV}$, i.e., the dimensionless couplings in \eqref{EM EFT} $\gamma,~\tilde{\gamma}$ are of order one.

However, the arguments following from black hole physics, known as Weak Gravity Conjecture (WGC) \cite{Arkani-Hamed:2006emk,2401.14449,1903.06239,2201.08380,2409.10003,Bastian:2020egp}, require the photon-graviton coupling corresponding to the $R F F$ term to be parametrically weaker than the one implied by the saturation of the unitarity bound \eqref{unitarity}. The original statement \cite{Arkani-Hamed:2006emk} follows from the requirement that all charged black holes must evaporate without forming naked singularities (Cosmic Censorship Conjecture). This implies that the objects with $M=Q$ in Planck units must have a horizon. In the presence of higher derivative EFT corrections, the extremality condition for the Reissner--Nordstr\"om (RN) black holes would be changed. This change must still keep $M=Q$ objects to be black holes, hence the condition on the EFT couplings \cite{Arkani-Hamed:2006emk, Bellazzini:2019xts,Henriksson:2022oeu},
\begin{equation}
\label{WGC}
    \left|\frac{\beta}{\Lambda_2^2\,M_P^2}\right|<\frac{2 \gamma}{\Lambda_{UV}^4}.
\end{equation}
If this bound is saturated, we have
\begin{equation}
\label{WGCchoice}\Lambda_2\simeq\frac{\Lambda_{UV}^2}{M_P}.
\end{equation}
It is interesting to mention that the same condition as \eqref{WGC} shows up in the other contexts, mainly related to RN black holes, such as Wald entropy formula \cite{Cheung:2018cwt} and stability of scalar quasinormal modes \cite{DiRusso:2025qpf}. However, there are no similarly looking results rigorously derived from analyticity and unitarity of scattering amplitudes, where the current state-of-the-art results \cite{Henriksson:2022oeu} still allow for significant violations of the WGC. 

In what follows, we will examine the implications of the described bounds on $\Lambda_2$ scale for the graviton production in the early Universe and for the high-frequency gravitational wave spectrum. We will perform computations for the two choices of this scale: the one corresponding to the saturation of unitarity \eqref{unitarity}, and, thus, leading to the maximal possible GW signal, and the one saturating the WGC bound \eqref{WGC}, which leads to less efficient graviton production.

\section{Production of gravitons during reheating}
\label{sec:graviton_prod}
\subsection{Differential decay rate}
\begin{figure}[htbp]
    \centering
    \includegraphics[width=1.\linewidth]{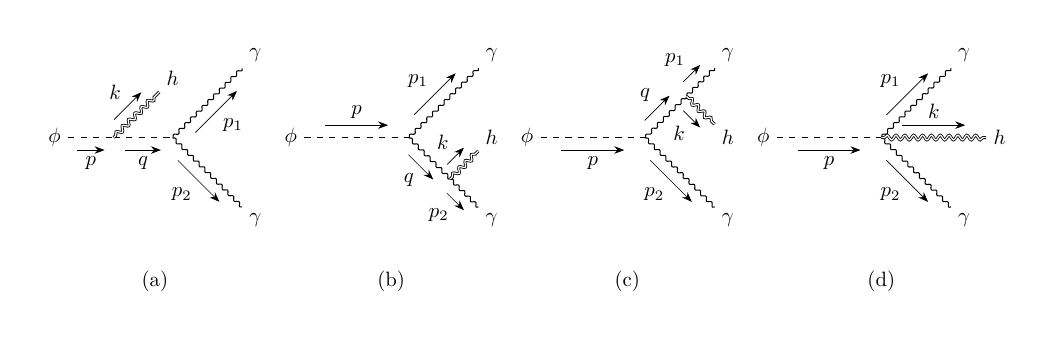}
    \caption{Feynman diagrams of photon bremsstrahlung and inflaton decay at the leading order. Here $\phi$ is an inflaton, $\gamma$ is a photon, $h$ is a graviton, and $p,q,p_1,p_2$ are momenta of the corresponding particles. The matrix element of panel (a) vanishes due to transverse and traceless condition for the graviton polarization tensor.}
    \label{fig:feynman diagram}
\end{figure}

During reheating, the gravitons are generated by the inflaton decay to two photons and a single graviton, see the corresponding Feynman diagrams in Figure \ref{fig:feynman diagram}.
We compute the matrix elements and the corresponding decay rates for the couplings following from the action \eqref{action} with the use of the xAct tensor algebra package for Wolfram Mathematica \cite{xActwebpage,0704.1756, 0807.0824, 0803.0862, 1308.3493}\footnote{We summarised all the derivations in the Mathematica file available online at \cite{xActJ}.}. The detailed results of the computation of the matrix elements and kinematic bounds on $q$ are given in the Appendix \ref{appendix: matrix elements}. The differential decay rate summed over all polarisation states of photons and gravitons reads,
\begin{align}\label{eq:differential decay rate 1}
    G(k)&=\frac{\partial \Gamma_{GW}}{\partial k}=\frac{1}{(2\pi)^3}\frac{1}{8m}\int^{q_{max}}_{q_{min}} dq{\left|\mathcal{M}(k,q)\right|}^2,
\end{align}
where $\left|\mathcal{M}(k,q)\right|^2$ stands for the sum of squared matrix elements over the polarisations, see the Appendix \ref{appendix: matrix elements} for the details. Here we provide the final result for the total differential decay rate as a function of the (comoving) energy of the graviton $k$,
\begin{align}\label{eq:differential decay rate}
G(k) &= \frac{m \alpha^2}{480 k (k - m)^3 M_P^2 \pi^3 \Lambda_1^2 \Lambda_2^4} \left[F_1(k) +F_2(k)\left(k - \frac{m}{2}\right) \log\left(1 - \frac{2k}{m}\right) \right].
\end{align}
Here
\begin{align*}
F_1(k) &= P_1(k) +P_2(k)+P_3(k),\quad F_2(k) = P_4(k) + P_5(k),
\end{align*}
and
\begin{align*}
	\left\{
		\begin{aligned}
			P_1(k) &= 64 k^4 (k - m)^3 m^2 \left(24 k^2 - 20 k m + 5 m^2\right) \beta^2\\
			P_2(k) &= -80 k^2 (k - m)^3 m^2 \beta \Lambda_2^2 \left(12 k^2 - 19 k m + 6 m^2\right) \\
			P_3(k) &=15 (k - m)\Lambda_2^4 \left(8 k^6 - 48 k^5 m + 76 k^4 m^2 - 60 k^3 m^3 + 25 k^2 m^4 - 6 k m^5 + m^6\right) \\
			P_4(k) &= 480 m^2k\Lambda_2^2  \beta (k - m)^2 (2 k - m) \left(k^2 - k m + m^2\right)\\
			P_5(k) &= 60mk\Lambda_2^4\left(4 k^4 - 12 k^3 m + 18 k^2 m^2 - 12 k m^3 + 3 m^4\right).
		\end{aligned}
	\right. 
\end{align*}
The graviton energy is bounded by $0<k<\frac{m}{2}$. The leading order expansion of the differential decay rate of $G(k)$ around $k=0$ is $\frac{1}{k}$, which corresponds to the IR divergence in the soft limit. Strictly speaking, this divergence requires a resummation of differential decay rates including the soft emission of many gravitons\cite{Barker:1969jk,Addazi:2019mjh,Ai:2025xla}. However, in the expanding Universe, this divergence can be eliminated by the fact that the emitted gravitons must be only sub-horizon gravitons. In this way, the horizon of the expanding Universe plays the role of the IR cutoff for the differential decay rate. This is a good approximation when the matrix element ${\cal M}$ is much smaller than 1, which appears to be the case for the scales considered in this work.

It is interesting to mention that the EFT couplings proportional to $\beta$ do not contribute to the IR divergence in the soft limit. The latter is fully determined by the photon-graviton vertex emerging from the kinetic term of the photon. Instead, they modify the differential decay rate at higher comoving momenta below the kinematic bound $k<m/2$. In the EFT expansion, the coupling parametrised by $\beta$ is the only possible contribution to the three-point vertex of graviton and photons, apart from the minimal coupling from the kinetic term. All other terms with more covariant derivatives can be reduced to four point interactions by perturbative field redefinitions. It means that there are no other contributions from EFT coupling of higher derivative terms which would affect the considered Feynman diagrams.

The term containing $\log{(1-\frac{2k}{m})}$ corresponds to the non-analytic behaviour of the differential decay rate in the collinear limit. It emerges because of the divergent on-shell limit of the internal propagator. However, in the case of graviton emission, this term is finite \cite{Weinberg:1965nx}. It can be understood from the fact that the three-point vertex of massless photons and gravitons also vanishes on shell, compensating the divergence coming from the propagator.\footnote{The computations in \cite{Barman:2023ymn} performed for the massive vector field still show the presence of such divergences in the formal limit $m\rightarrow 0$. However, for vector fields, the formal massless limit cannot be taken because the number of degrees of freedom in the massless case is less than for non-zero mass.}

\subsection{Boltzmann equation}
In order to connect the differential decay rate \eqref{eq:differential decay rate} with the cosmological observables, we need to compute the spectrum of gravitons produced during reheating and its subsequent evolution in the expanding Universe. The latter can be described by the Boltzmann equation.

At time $t$, within the energy interval $k(t)\sim k(t)+\dd{k(t)}$, gravitons have number distribution $\dv{N_{h}}{k(t)}$. Thus, we have the following energy density distribution at interval $k(t)\sim k(t)+\dd{k(t)}$,
\begin{align}
    \dd{\rho}_{GW}(t)=k(t)\frac{1}{V(t)}\dv{N_{h}}{k(t)}\dd{k(t)}=k(t)\frac{1}{V(t)}\dv{N_{h}}{k}\dd{k}.
\end{align}
Here $V(t)$ is the comoving volume at time $t$, $k(t)=k/a(t)$ is a physical momentum, corresponding to the conformal momentum $k$ of the graviton. The latter can be related to the energy of the graviton at present if we take the scale factor $a_0=1$. Thus, we can write
\begin{align}
    \dv{\rho_{GW}(t)}{k}=k(t)\dv{k}\left[\frac{N_h}{V(t)}\right]=\frac{k}{a^4(t)}\dv{k}\left[\frac{N_h}{V_0}\right].
\end{align}
Performing a time derivative on both sides of the equation, we obtain 
\begin{align}\label{eq:appendix b 1}
    \dv{\dot{\rho}_{GW}(t)}{k}+4H(t)\dv{\rho_{GW}(t)}{k}=\frac{k(t)}{V(t)}\dv{k}\dv{t}N_h.
\end{align}

In our case, we have one graviton produced by one inflaton decay, such that 
\begin{align}
    \dv{N_h}{t}=-\dv{N_{\phi}}{t}=N_{\phi}\left(\frac{-1}{N_{\phi}}\dv{N_{\phi}}{t}\right)=N_{\phi}\Gamma_{\phi},
\end{align} 
where $N_{\phi}$ is the number of inflaton and $\Gamma_{\phi}$ is the decay rate of inflaton.
Then equation \eqref{eq:appendix b 1} can be written as 
\begin{align*}
    \dot{\rho}_{GW}(t)+4H(t)\rho_{GW}(t)
    =&\int \dd{k}\frac{k(t)}{V(t)}N_\phi\dv{\Gamma_{\phi}}{k}\\
    =& \int \dd{k} \frac{k(t)}{E_{\phi}}\frac{E_{\phi}N_{\phi}}{V(t)}\dv{\Gamma_{\phi}}{k}\\
    =& \int \dd{k} \frac{k(t)}{E_{\phi}}\frac{E_{\phi}N_{\phi}}{V(t)}\dv{\Gamma_{\phi}}{k(t)}\dv{k(t)}{k}\\
    =& \int \dd{k} \frac{k(t)}{E_{\phi}}\rho_{\phi}(t)G[k(t)]\frac{1}{a(t)},
\end{align*}
where $E_{\phi}$ is the energy of an inflaton. 

In our case, given that we assumed that reheating can be described as the decay of the condensate of $\phi$-particles with zero momenta, $E_{\phi}=m$, therefore we obtain, 
\begin{align}\label{eq:Boltzmann equation for graviton}
    \dv{t}\rho_{GW}(t)+4H(t) \rho_{GW}(t)=\int \dd{k'} \frac{k'(t)}{m} \rho_{\phi}(t) G[k'(t)]\frac{1}{a(t)},
\end{align}
which is the Boltzmann equation for the energy density of gravitons, in the case of bremsstrahlung during reheating.

\subsection{Gravitational wave spectrum}

Having the Boltzmann equation \eqref{eq:Boltzmann equation for graviton}, we can connect the energy density $\rho_{GW}(t)$ with the late-time observable $\dv{\Omega_{GW}}{\log k }$ which the spectrum of GWs at the present time expressed in terms of the relative energy fraction stored in GWs.

\begin{figure}[htbp]
    \centering
    \includegraphics[width=0.5\linewidth]{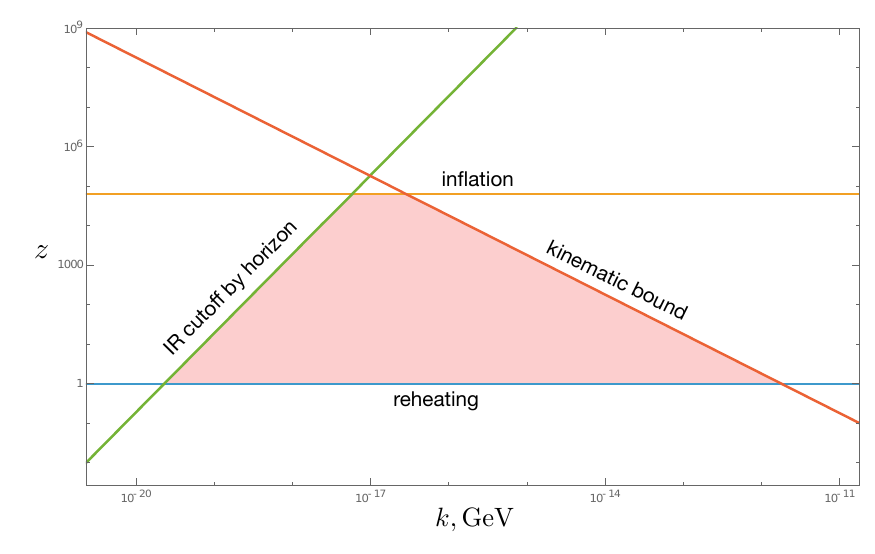}
    \caption{Integration limits on $z$. Here we set $\alpha=\beta=1$, $H_{inf}=10^{12}$ GeV, $g_{reh}=106.75$, $\Lambda_1=\Lambda_{UV}$, $\Lambda_2=\Lambda_{UV}^3/M_P^2$, $\Lambda_{UV}=10^{16.5}$ GeV, $m=10^{13}$ GeV.}\label{fig:integral limits}
\end{figure}

From \eqref{eq:Boltzmann equation for graviton}, we have 
\begin{align*}
    \dv{t}\left[\dv{\rho_{GW}(t)}{k}\right]+4H(t)\dv{\rho_{GW}(t)}{k}=\frac{k(t)}{m}\rho_{\phi}(t)G[k(t)]\frac{1}{a(t)},
\end{align*}
If we multiply both sides of equation by $a^4(t)$, we obtain
\begin{align*}
        \dv{t}\left[a^{4}(t)\dv{\rho_{GW}(t)}{k}\right]
        &=a^4(t)\frac{k(t)}{m}\rho_{\phi}(t)G\left[k(t)\right]\frac{1}{a(t)}\\
        &=a^4(t)k(t)n_{\phi}(t)G\left[k(t)\right]\frac{1}{a(t)}\\
        &=kn_{\phi}(t)a^2(t) G\left[k\frac{1}{a(t)}\right].
\end{align*}
Furthermore, the left hand side of this equation can be written as $\dv{t}\left[a^4(t)\dv{\rho_{GW}(t)}{k}\right]=\dv{t}\left[\dv{\rho_{GW,0}}{k}\right]$, since $\rho_{GW}(t)\propto a^{-4}(t)$. Then the equation above becomes,
\begin{align*}
     \dv{t}\left[\dv{\rho_{GW,0}}{k}\right]=kn_{\phi}(t)a^2(t) G\left[k\frac{1}{a(t)}\right].
\end{align*}
 Integrating it over time $t$ we obtain 
\begin{align*}
    \dv{\rho_{GW,0}}{k}=\int_{t_1}^{t_2} \dd{t} k n_{\phi}(t) a^2(t) G\left[k\frac{1}{a(t)}\right],
\end{align*}
which is the GW energy density distribution, and $t_1$, $t_2$ are the moments when the inflaton decay began and ended, respectively. Then we define the GW spectrum as,
\begin{align}\label{eq: energy density distribution 1}
    \frac{d \Omega_{\text{GW}}}{d \log k} &\equiv \frac{k}{\rho_0} \frac{d \rho_{\text{GW,0}}}{d k} = \frac{k}{\rho_0} \int_{t_1}^{t_2} \dd{t} k n_{\phi}(t) a^2(t) G\left[k \frac{1}{a(t)}\right], 
\end{align}
where $\rho_0=3M_P^2H_{0}^2$ is the current energy density. Since we have
\begin{align*}
    n_{\phi}(t)&\equiv\frac{N_{\phi}(t)}{V(t)}=\left(N_{reh}e^{-\Gamma t}\right)\left(\frac{a_{reh}^3}{a^3(t)}\frac{1}{V_{reh}}\right)\\
    &=\frac{a_{reh}^3}{a^3(t)}\frac{1}{m}\left(m\frac{N_{reh}}{V_{reh}}\right)e^{-\Gamma t}\\
    &=\left(\frac{a_{reh}}{a(t)}\right)^3\frac{\rho_{reh}}{m}e^{-\Gamma t},
\end{align*}
where $\Gamma$ is the total decay rate of the inflaton and $N_{reh}$ is the total number of inflaton particles at the time when the inflaton decay began. Then the equation \eqref{eq: energy density distribution 1} can be written as 
\begin{align}\label{eq: energy density distribution 2} 
    \dv{\Omega_{GW}}{\log k}&=\frac{k}{\rho_0} \int_{t_1}^{t_2} \dd{t} k a^2(t) \left(\frac{a_{reh}}{a(t)}\right)^3\frac{\rho_{reh}}{m}e^{-\Gamma t}G\left[k \frac{1}{a(t)}\right]\nonumber \\
    &=\frac{k^2}{m}\frac{\rho_{reh}}{\rho_0}a_{reh}^2\int_{t_1}^{t_2}\dd{t} \frac{a_{reh}}{a(t)} e^{-\Gamma t}G\left[k \frac{1}{a(t)}\right].
    \end{align}
We can define an integration variable $z=\frac{a_{reh}}{a(t)}$. At the matter-dominated stage during the period of inflaton oscillations, we have $a(t)=ct^{\frac{2}{3}}$, thus, we can write,
\begin{align*}
     t=\left[\frac{a(t)}{c}\right]^{\frac{3}{2}},\enspace H(t)=\frac{\dot{a}(t)}{a(t)}=\frac{2}{3}t^{-1}=\frac{2}{3} \left[\frac{c}{a(t)}\right]^{\frac{3}{2}}.
\end{align*}
Here $c$ is a constant relating the scale factor to time. Thus,
\begin{align*}
    \left\{
        \begin{aligned}
            &\exp(-\Gamma t)=\exp(-H_{reh}\left[\frac{a(t)}{c}\right]^{\frac{3}{2}})=\exp(-\frac{2}{3}z^{-\frac{3}{2}})\\
            &\dd{t}=\frac{-dz}{Hz}=\frac{-\dd{z}}{H_{reh}z^{\frac{5}{2}}}\\
            &G\left[k\frac{1}{a(t)}\right]=G\left[kz\frac{1}{a_{reh}}\right]
        \end{aligned}
    \right..
\end{align*}
The first equation here comes from the condition determining the moment of reheating $\Gamma\simeq H(t)$. Therefore, the equation \eqref{eq: energy density distribution 2} can be written as 
\begin{align}\label{eq:energy density distribution 3}
    \dv{\Omega_{GW}}{\log k}
    =\frac{k^2}{mH_{reh}}a_{reh}^2\frac{\rho_{reh}}{\rho_0}\int_{z_{min}}^{z_{max}}\dd{z}z^{-\frac{3}{2}}\exp(-\frac{2}{3}z^{-\frac{3}{2}})G\left[kz\frac{1}{a_{reh}}\right].
\end{align}
Now we need to determine the integral limits $z_{min}$ and $z_{max}$ referring to the time duration when the GW signal has been generating. We have the following constraints:
	\begin{itemize}
		\item As we are computing GWs emitted between inflation and reheating, we require $1<z<\frac{a_{inf}}{a_{reh}}$, or
        \begin{equation}
            {1<z <z_1=  \left( \frac{3 H_{inf}^2 M_P^2}{g_{reh} T_{reh}^4 \, \pi^2/30} \right)^{1/3}}.
        \end{equation}
		\item Here we implement the assumption that all generated gravitons are sub-horizon, 
\begin{equation}
    k\gg a(z)H(z)\Rightarrow {z < z_2 = \frac{k^2 a_0^2}{a_{reh}^2 \, H_{reh}^2} \,.}
\end{equation}
        This requirement provides a natural IR cutoff for the IR divergence in the soft limit of the differential decay rate. At the horizon scale, the graviton production cannot be described by the Feynman rules derived for flat spacetime. An accurate computation requires this process to be considered in the expanding Universe, which is left for future work.\footnote{The absence of superhorizon gravitons can be physically motivated by causality. As the bremsstrahlung is a localized process, one cannot expect to produce the effects at length scales larger than the size of the horizon.} 
		\item Kinematic upper bound on the energy of the graviton,
        \begin{equation}
            k<\frac{m}{2}\Rightarrow z<\frac{m}{2}\frac{a_{reh}}{a_0}.
        \end{equation}
	\end{itemize}
All the limits described here are shown in Figure \ref{fig:integral limits}. It illustrates that all the mentioned constraints are essential and must be taken into account for computing the GW energy spectrum.
The GW spectrum at high frequencies is usually expressed in terms of the characteristic strain \cite{Maggiore_2000},
	\begin{align}\label{eq:characteristic strain}
		h_c(k)=\sqrt{\frac{6  H_0^2}{k^2}\dv{\Omega_{GW}}{\log k }},
	\end{align}
where $k=2\pi f$. Then, the equations  \eqref{eq:differential decay rate} \eqref{eq:energy density distribution 3}, and \eqref{eq:characteristic strain} allow to obtain the predictions for the GW spectrum, thus connecting the EFT description of reheating with the potentially observable high-frequency GW signals. 
\subsection{Plots of GW spectrum}
We performed computations of the GW spectrum for a certain range of EFT parameters describing a model of inflation and reheating. In total, given the assumptions about the natural choice of EFT scales described in Section 3, the setup is fully determined by the inflaton mass $m$, UV-cutoff of the theory $\Lambda_{UV}$, and the scale $\Lambda_2$, (in a combination $\beta/\Lambda_2^2$) suppressing the $R F F$ coupling between the photon and graviton. 

\begin{figure}[H]  
    \label{GWplots}
    \centering
\includegraphics[width=0.45\textwidth]{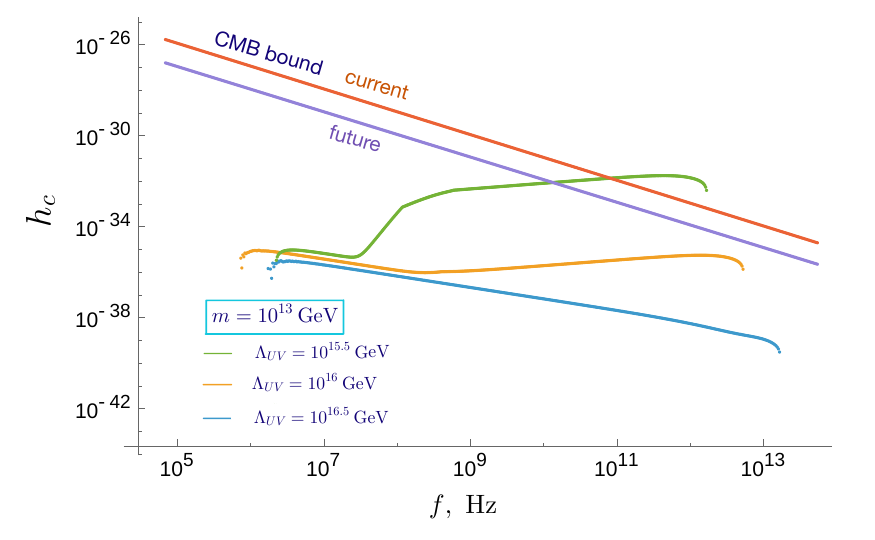} 
\includegraphics[width=0.45\textwidth]{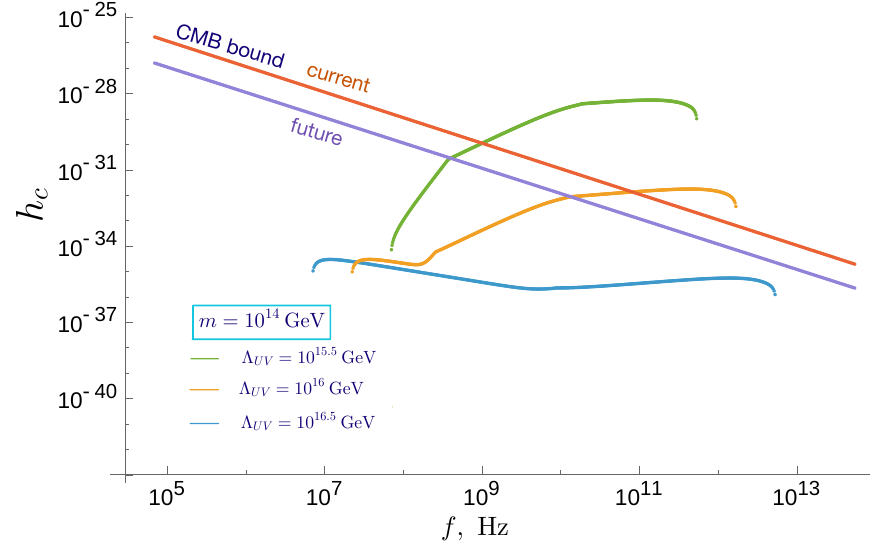}
\includegraphics[width=0.45\textwidth]{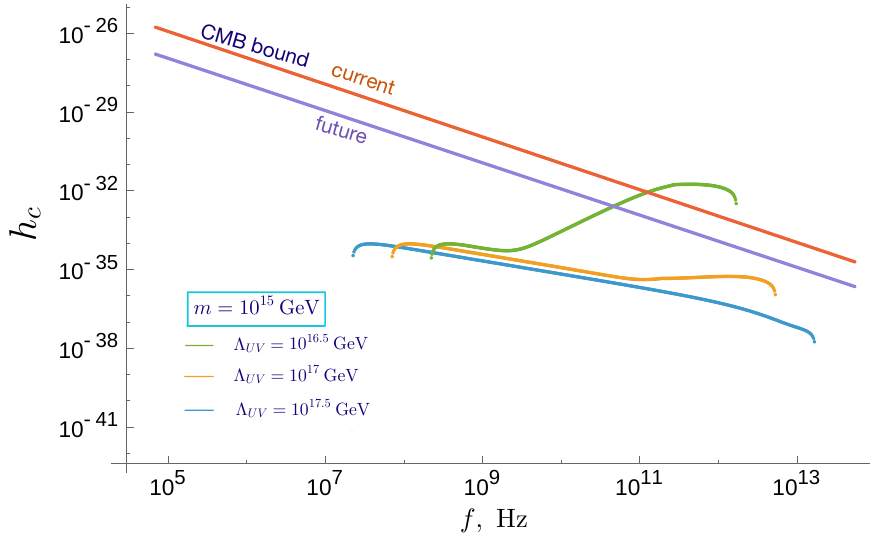} 
\includegraphics[width=0.45\textwidth]{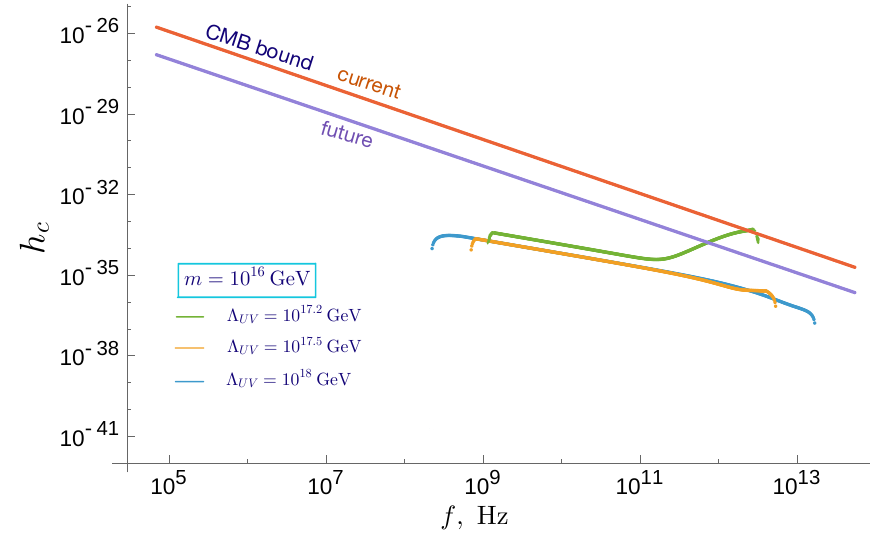}
    \caption{Examples of high frequency gravitational wave signal for the reheating scenarios determined by the choice $\Lambda_1=\Lambda_{UV}$, $\Lambda_2=\Lambda_{UV}^3/M_P^2$. The red and purple lines show the current CMB bound \cite{Planck:2018vyg} on the number of relativistic degrees of freedom and the bound that can be potentially obtained in future observations, respectively.}  
    \label{fig: bound}  
\end{figure}

Figure \ref{GWplots} shows the expected GW signals, given the choice of $\Lambda_2=\Lambda_{UV}^3/M_P^2$, as in \eqref{unitarity}. These results should be referred to as the maximal possible values of characteristic strain in an EFT consistent with unitarity. We see that for the scale $\Lambda_{UV}\sim 10^{16}$ GeV, these signals can be large and exceed the existing CMB bound, thus providing a constraint on the UV cutoff which will be discussed in detail in the next Section \ref{sec:constraints}. 

The signals obtained for different values of the model parameters have the following common features in their shape:
\begin{itemize}
    \item The frequency range is typically $10^7$ - $10^{13}$ Hz. It is bounded by the IR cutoff, as an implication of the fact that we excluded production of the superhorizon gravitons, see Figure \ref{fig:integral limits}. The upper bound is a kinematic constraint on the comoving momentum of the produced graviton. 
    \item The IR part of the spectrum does not strongly depend on the scale $\Lambda_2$, as it is dominated by the $1/k$ part of the differential decay rate, which is not modified by $R F F$ coupling. In terms of characteristic strain, it is growing for lower frequencies until the horizon cutoff comes into the game. The same features were obtained in the literature \cite{Choi:2024ilx,Hu:2024awd,Xu:2024fjl,Inui:2024wgj,Jiang:2024akb,Strumia:2025dfn,Bernal:2025lxp,Nakayama:2025xkn,Barman:2023ymn,Huang:2019lgd} for the case of minimal Planck-suppressed coupling between graviton and matter. 
    \item The spectrum has a rising part in the UV regime, which is determined by the coupling $R F F$. Such a feature is absent in the case of minimal coupling \cite{Barman:2023ymn}, as well as in the case of inflaton decay to scalars in the EFT framework \cite{Tokareva:2023mrt}. This is related to the higher powers of $k$ in the differential decay rate, provided that the decay responsible for reheating is also due to the higher-derivative operator.
\end{itemize}
\begin{figure}[H] \label{spectra} 
    \centering
\includegraphics[width=0.9\textwidth]{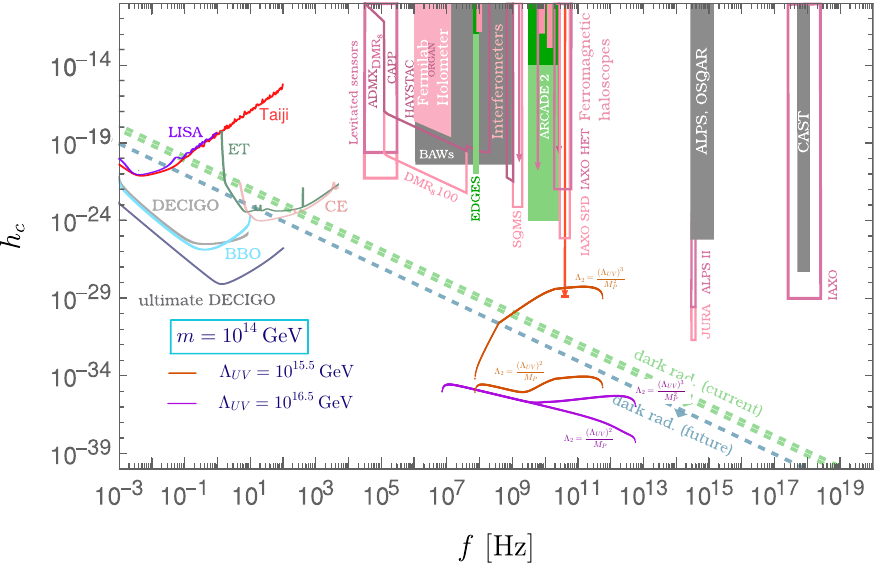} 
    \caption{The dependence of the GW signal on the parametric choice of the scale $\Lambda_2$. }  
    \label{fig: bound}  
\end{figure}
Figure \ref{spectra} shows the position of the signals from reheating in a full range of frequencies which can be potentially probed in future experiments. The plot emphasises the difference between the two choices of the $R F F$ coupling - the strongest one still allowed by unitarity constraints ($\Lambda_2\simeq \Lambda_{UV}^3/M_P^2$), and the one which respects the black hole Weak Gravity Conjecture \eqref{WGC} ($\Lambda_2\simeq \Lambda_{UV}^2/M_P$). The latter implies much weaker signals in general, keeping them safe from violating the observational constraints. However, in any case, the detection of gravitational waves at such high frequencies is challenging because so far, all future proposals for detecting the high-frequency gravitons cannot reach the values of the strain below the CMB bound. At the same time, the wide signals with higher values of the strain are ruled out by the CMB bound. \footnote{The CMB bound is a constraint on the spectrum integrated over all the frequencies, so, in principle, a narrow peak can still go higher than the green dashed line in Figure \ref{spectra}. This line represents the constraint for the flat spectrum at lower frequencies; however, the broad signals going over this line are also ruled out.} 

In the setup under consideration, given that the coupling between photon and gravity is larger than in the case of Planck scale suppression, there is a chance that the detection methods based on graviton-photon conversion in strong magnetic fields would be more sensitive to high-frequency gravitons than is usually assumed. In this case, the described reheating scenarios can still be probed; however, the estimates of their sensitivity must be recomputed with the inclusion of the EFT coupling $R F F$. This could also be a very interesting possibility to probe the structure of the EFT of photons and gravity, as in general one can also expect the graviton production by the thermal plasma after reheating, a contribution which is unavoidable and independent of the concrete reheating scenario. The latter was computed only for the minimal coupling between the SM fields and gravity \cite{Ghiglieri:2020mhm,Ghiglieri:2024ghm}, however, the presence of the EFT terms should also enhance such signals.

\section{Constraints on the UV cutoff scale}
\label{sec:constraints}
As can be seen from the plots of the gravitational wave signals (Figure \ref{GWplots}) for different parameters, such as the inflaton mass and UV cutoff of the theory, some of these signals exceed even the current CMB bound. It means that such scenarios are already ruled out, and the future constraints are capable of improving the bound. 

\begin{figure}[H]
\label{cutoff constraint}
    \centering
\includegraphics[width=0.45\textwidth]{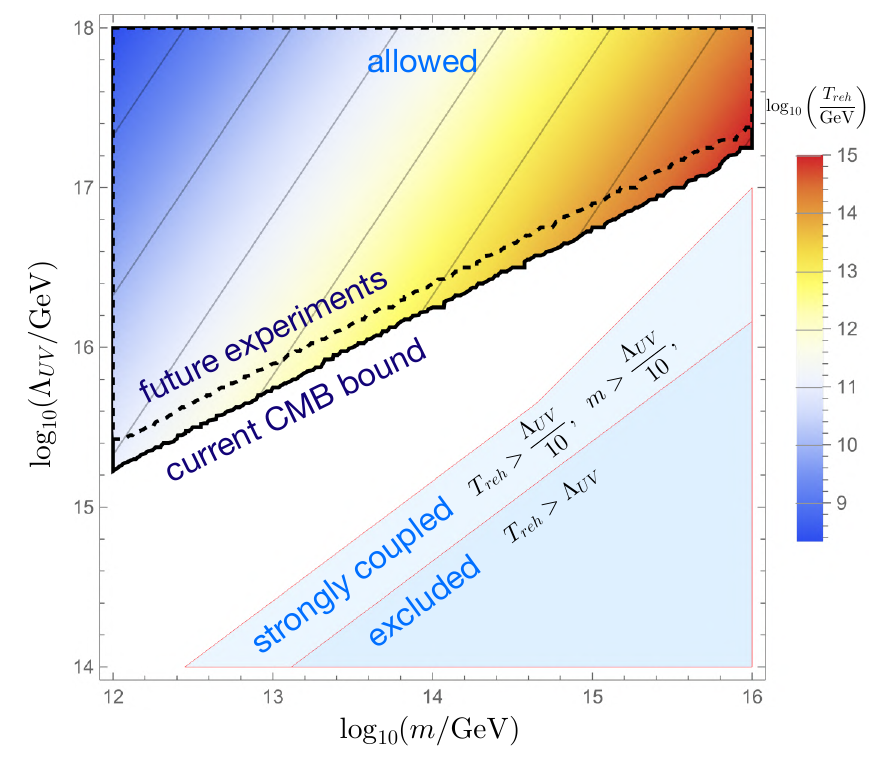} 
\includegraphics[width=0.45\textwidth]{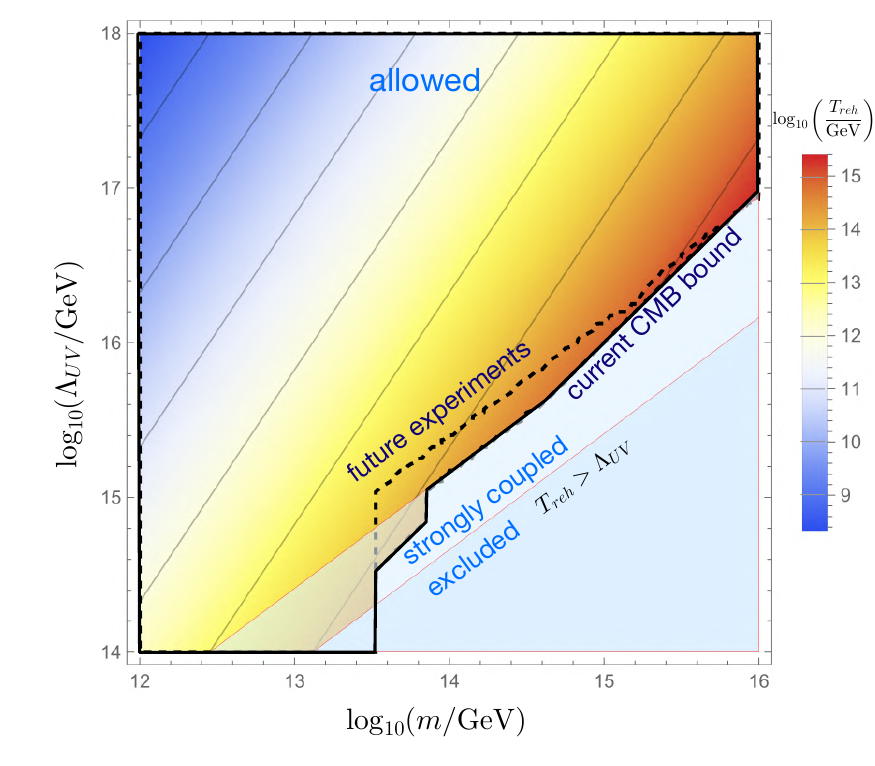}
    \caption{Constraints imposed by the current CMB data on the UV cutoff of gravitational EFT and inflaton mass. The values in the axis are given in $\log_{10}(\Lambda_{UV}/{\rm GeV})$ and $\log_{10}(m/{\rm GeV})$, respectively. The values of reheating temperature are shown in color.}  
    \label{fig: bound}  
\end{figure}

If we take it as an assumption that the value of $R F F$ coupling is close to the unitarity bound, $\Lambda_2\simeq\Lambda_{UV}^3/M_P^2$, in order to avoid the graviton overproduction, we must require the scale $\Lambda_{UV}$ to be high enough. Given the choice of $\Lambda_2$ is made, the model is fully parametrised by the inflaton mass $m$ and UV cutoff scale $\Lambda_{UV}$. In this case, the requirement that the GW signal never exceeds the CMB bound imposes a constraint on the parameters $m$ and $\Lambda_{UV}$. The left plot of Figure \ref{cutoff constraint} represents the allowed region for the theories with $\Lambda_2=\Lambda_{UV}^3/M_P^2$, and the right plot is for theories satisfying the WGC, $\Lambda_2=\Lambda_{UV}^2/M_P$ (we take $\beta=1$; we found that the sign of $\beta$ doesn't affect the results). A part of the parameter space (light blue regions) is forbidden by the self-consistency of the EFT description of reheating, which requires all the physical scales, such as inflaton mass and reheating temperature, to be less than $\Lambda_{UV}$. We can see that the allowed parameter region is very sensitive to the parametric scalings of the EFT couplings. In particular, the theories violating WGC are required to have a very high unitarity breaking scale $\Lambda_{UV}>10^{15}$ GeV for the reasonable choice of parameters for the large field inflation. However, if the WGC requirement suppresses the photon-graviton interaction, the constraints imposed by the CMB bound forbid only strongly coupled scenarios. The future constraints on dark radiation can provide only a little improvement for the excluded range of parameters.

Does the WGC requirement imply only very low gravitational wave signals? In fact, the black hole arguments in favor of the WGC constraint \eqref{WGC} are strongly related to the global $U(1)$ charge and RN solutions; thus, they can be applied only for photons. For the other gauge bosons (gluons, $W$- and $Z$-bosons), there is only a unitarity constraint still allowing for the scaling $\Lambda_2\sim \Lambda_{UV}^3/M_P^2$. Thus, our results can be straightforwardly extended to the reheating scenarios with inflaton decay to the other SM gauge fields, where there are no WGC-type arguments parametrically suppressing the coupling between graviton and gauge fields. 

In the EFT setup used in this paper one could also expect the other couplings potentially contributing to the graviton production in the early Universe. However, as the leading signal is related to the emission of a single graviton, the vertices responsible for such an emission emerge only from the kinetic term of photon and $RFF$ term (see, for example. \cite{Ruhdorfer:2019qmk,Henriksson:2022oeu} for a construction of the full set of couplings). The other terms with more derivatives would only produce vertices with two or more gravitons, which are additionally suppressed by the Planck scale. For this reason, the obtained results for graviton production are solid and would not be affected by higher-order EFT corrections.

\section{Conclusions and outlook}
\label{sec:conclusions}

In this work, we investigated the production of high-frequency gravitational waves during the reheating phase of the early Universe, focusing on the EFT framework that describes the interactions between the inflaton, gravitons, and photons. We considered an EFT construction where the inflaton at preheating stage decays mainly into photons through a dimension-5 operator, $\phi F_{\mu\nu}F^{\mu\nu}$, and examined the subsequent production of gravitons via bremsstrahlung, mediated by the dimension-6 operator $R_{\mu\nu\rho\sigma}F^{\mu\nu}F^{\rho\sigma}$. By computing the differential decay rate of the inflaton into two photons and a graviton, we derived the spectrum of the resulting GW signal. Our calculations revealed an IR divergence in a soft limit of the differential decay rate, as well as a distinctive non-analytic term, $\log(1 - 2k/m)$. The latter arises in the collinear limit of the process and is absent in the case of inflaton decay to scalars.

Using the Boltzmann equation, we connected the differential decay rate to the observable GW spectrum, expressed in terms of the characteristic strain $h_c$. Our results demonstrated that the GW spectrum is remarkably sensitive to the relative choice of the scales in the EFT action $\Lambda_1$ and $\Lambda_2$. Taking $\Lambda_1=\Lambda_{UV}$, we considered two distinctive possibilities for $\Lambda_2$: the one corresponding to the saturation of the unitarity bound ($\Lambda_2 \simeq \Lambda_{UV}^3 / M_P^2$) and the other one expected from saturation of the Weak Gravity Conjecture ($\Lambda_2 \simeq \Lambda_{UV}^2 / M_P$). The former leads to a stronger GW signal, while the latter suppresses the signal due to the weaker photon-graviton coupling.

By comparing the predicted GW signals with the current CMB bound on dark radiation, we found that the scenarios with $\Lambda_2 \simeq \Lambda_{UV}^3 / M_P^2$ can exceed the observational constraints for certain values of the inflaton mass and UV cutoff. This imposes a lower bound on the UV cutoff scale, $\Lambda_{UV} \gtrsim 10^{15}$~GeV, for inflaton masses typical of large-field inflation models ($m \gtrsim 10^{12}$~GeV). In contrast, theories respecting the WGC are less constrained, as the weaker photon-graviton coupling results in GW signals that are more consistent with observational limits.

Our findings highlight the importance of high-frequency GW signals as a potential probe of the EFT structure of gravity and inflation. While direct detection of such high-frequency GWs remains challenging with current and near-future experiments, the CMB bound on dark radiation provides a powerful indirect constraint. Future improvements in CMB measurements could further tighten these constraints and potentially open a new window into the physics of reheating and the UV completion of gravity.

Although this work was mainly focused on the case of the photon decay channel responsible for reheating, the results can be straightforwardly generalized to the other SM gauge bosons and to the EFT operators describing matter coupled to gravity. A very promising future direction is related to the development and the implications of the EFT bootstrap constraints on theories coupled to gravity, extending the recent results on the species bound \cite{Caron-Huot:2024lbf} or lower bound on the scalar self-coupling. Such methods are also capable of constraining the ratio $\Lambda_{UV}/M_P$. 

It is still unclear if there is a way to derive a similar scaling for the $R F F$ coupling, as it is imposed by the black hole WGC arguments from dispersion relations or other bootstrap methods in flat spacetime. It is also unclear whether the constraints beyond flat spacetime, such as black hole WGC, can be derived at all from the S-matrix. Although this work highlights this difference in theoretical constraints and emphasises that the violations of the WGC can be constrained from cosmological observables, it is unclear whether theories violating WGC can be ruled out by more solid theoretical arguments, such as unitarity and causality. The methods developed in \cite{CarrilloGonzalez:2022fwg,CarrilloGonzalez:2023cbf} allow for the causality probes beyond the flat spacetime \cite{CarrilloGonzalez:2023emp,CarrilloGonzalez:2025fqq}. Extending such constraints beyond the flat backgrounds can be a very promising direction for future studies.

\subsection*{Acknowledgements}

The authors are grateful to Ivano Basile, Marianna Carrillo-Gonzalez, Alexey Koshelev, Xunjie Xu for their valuable comments and discussions. 
The work of A.~T. was supported by the National Natural Science Foundation of China (NSFC) under Grant No. 12547104 and No. 12505091. 

\appendix
\section{Conventions}
In this work, we use the natural units $\hbar=c=1$, metric $(1,-1,-1,-1)$. We define the metric fluctuation around the flat spacetime as 
\begin{align}
    g_{\mu\nu}=\eta_{\mu\nu}+\frac{2}{M_P}h_{\mu\nu},\quad g^{\mu\nu}=\eta^{\mu\nu}-\frac{2}{M_P}h^{\mu\nu},
\end{align}
where $M_P=\frac{1}{\sqrt{8\pi G}}$ is the reduced Planck mass. 
We choose de Donder gauge for the graviton $\partial_{\mu}h^{\mu}{}_{\nu}-\frac{1}{2}\partial_{\nu}h^{\lambda}{}_{\lambda}=0$.
In this gauge choice metric perturbations are transverse and traceless, i.e. ${\partial_{\mu}h^{\mu}{}_{\nu}=h^{\mu}{}_{\mu}=0}$.
For the photon, we use the Lorentz gauge $\partial_{\mu}A^{\mu}=0$.
Thus, the scalar propagator is 
\begin{align}
    \frac{i}{p^2-m^2+i\epsilon},
\end{align}
and the photon propagator is 
\begin{align}
    \frac{-i g^{\mu\nu}}{p^2+i\epsilon}.
\end{align}

\section{Matrix elements}\label{appendix: matrix elements}

To evaluate the value of the total matrix element squared $\left|\mathcal{M}(k,q)\right|^2$ obtained from the Feynamn diagrams \ref{fig:feynman diagram}, we choose such a frame that 
	\begin{align*}
        &\begin{aligned}
             p^{\mu}=\begin{pmatrix}
            m\\0\\0\\0
        \end{pmatrix},
        k^{\mu}=\begin{pmatrix}
            k\\0\\0\\k
        \end{pmatrix},
        p_1^{\mu}=\begin{pmatrix}
            q\\q \cos(\theta)\\0\\q \sin(\theta)
        \end{pmatrix},
        p_2^{\mu}=\begin{pmatrix}
            m-k-q\\-q\cos(\theta)\\0\\-k-q\sin(\theta)
        \end{pmatrix},
	\end{aligned}\\
	&\begin{aligned}
            &\epsilon^{(+)}_{\mu}(p_1)=\begin{pmatrix}
            0&0&1&0
        \end{pmatrix}
        ,\epsilon_{\mu}^{(-)}(p_1)=\begin{pmatrix}
            0&-\sin(\theta)&0&\cos(\theta)
        \end{pmatrix},\\
        &\epsilon^{(+)}_{\mu}(p_2)=\begin{pmatrix}
            0&0&1&0
        \end{pmatrix},\epsilon_{\mu}^{(-)}(p_2)=\frac{1}{m-k-q}\begin{pmatrix}
            0&k+q\sin(\theta)&0&-q\cos(\theta)
        \end{pmatrix},\\
        &\epsilon^{(+)}_{\mu\nu}(k) = \begin{pmatrix}
            0 & 0 & 0 & 0 \\
            0 & \frac{1}{2} & \frac{i}{2} & 0 \\
            0 & \frac{i}{2} & -\frac{1}{2} & 0 \\
            0 & 0 & 0 & 0
            \end{pmatrix}           
            , \quad
            \epsilon_{\mu\nu}^{(-)}(k)= \begin{pmatrix}
            0 & 0 & 0 & 0 \\
            0 & \frac{1}{2} & -\frac{i}{2} & 0 \\
            0 & -\frac{i}{2} & -\frac{1}{2} & 0 \\
            0 & 0 & 0 & 0
            \end{pmatrix},\\
            \end{aligned}
    \end{align*}
    where $p^{\mu}$ is the momentum of inflaton, $k^{\mu}$ is the momentum of the outgoing graviton, $p_1^{\mu}$ and $p_2^{\mu}$ are the momenta of the outgoing photons. $\epsilon^{(\pm)}$s are the polarization tensors of the corresponding particles.
    From $p_{2}^{2}=0$, we have 
    \begin{align}
        &\sin(\theta)=\frac{-2 k m+2 k q+m^2-2 m q}{2 k q}.
    \end{align}
    And since $-1<\sin(\theta)<1$, one may obtain the integration limits for the equation \eqref{eq:differential decay rate 1} as
    \begin{align}
         0<\frac{m}{2}-k<q<\frac{m}{2},
    \end{align}
    which also offers a kinetic bound for graviton energy: $0<k<\frac{m}{2}$. 
    
   Summing over all diagrams presented in the Figure \ref{fig:feynman diagram} under the specific choice of polarizations, one may obtain the matrix element squared $\left|\mathcal{M}_{ijk}\right|^2$ as
    \begin{align*}
        \left|\mathcal{M}_{+++}\right|^2&=\frac{m^2 \alpha^2 }{k^2 M_P^2 \Lambda_1^2 \Lambda_2^4}\left( 4 k m \left( 2 k^2 + (m - 2 q)^2 + k (-3 m + 4 q) \right) \beta - \left( 2 k^2 - 2 k m + m^2 \right) \Lambda_2^2 \right)^2,\\
        \left|\mathcal{M}_{++-}\right|^2&=\frac{m^2 \alpha^2 }{k^2 M_P^2 \Lambda_1^2 \Lambda_2^4}\left( 4 k m \left( 2 k^2 + (m - 2 q)^2 + k (-3 m + 4 q) \right) \beta - \left( 2 k^2 - 2 k m + m^2 \right) \Lambda_2^2 \right)^2,\\
        \left|\mathcal{M}_{+-+}\right|^2&=\frac{m^2 \alpha^2}{M_P^2 \Lambda_1^2}\left( -2k + 2m + \frac{m(-2k + m)}{k - m + q} - \frac{4m \left( -2k^2 + km - 2(m - 2q)q \right)\beta}{\Lambda_2^2} \right)^2,\\
        \left|\mathcal{M}_{+--}\right|^2&=\frac{m^2 \alpha^2}{M_P^2 \Lambda_1^2}\left( 2\left(k - m\right) + \frac{(2k - m)m}{k - m + q} + \frac{4m \left( -2k^2 + km - 2(m - 2q)q \right)\beta}{\Lambda_2^2} \right)^2,\\
        \left|\mathcal{M}_{-++}\right|^2&=\frac{m^2 \alpha^2 }{M_P^2 q^2 \Lambda_1^2 \Lambda_2^4}\left( -8 k^2 m q \beta + 4 k m (5m - 8q) q \beta + 2k (m - q) \Lambda_2^2 - m (m - 2q) \left( 8(m - q) q \beta + \Lambda_2^2 \right) \right)^2,\\
        \left|\mathcal{M}_{-+-}\right|^2&=\frac{m^2 \alpha^2 }{M_P^2 q^2 \Lambda_1^2 \Lambda_2^4}\left( 8 k^2 m q \beta + 4 k m q (-5m + 8q) \beta + 2k (-m + q) \Lambda_2^2 + m(m - 2q)\left( 8(m - q) q \beta + \Lambda_2^2 \right) \right)^2,\\
        \left|\mathcal{M}_{--+}\right|^2&=\frac{m^2 \alpha^2 \left( B_1 + B_2 \right)^2}{k^2 M_P^2 q^2 (k - m + q)^2 \Lambda_1^2 \Lambda_2^4},\\
        \left|\mathcal{M}_{---}\right|^2&=\frac{m^2 \alpha^2 \left( B_1 + B_2 \right)^2}{k^2 M_P^2 q^2 (k - m + q)^2 \Lambda_1^2 \Lambda_2^4},
    \end{align*}
    and
    \begin{align*}
        B_1 &= 4 k m \left[ 2 k^2 + k(m - 4q) - (m - 2q)^2 \right] q (k - m + q) \beta,\\
        B_2 &= \left[ k^2 m (-2k + m) + (k - m)(2k^2 - 2km + m^2)q + (2k^2 - 2km + m^2)q^2 \right] \Lambda_2^2.    
    \end{align*}
 Here, polarization indices $\alpha,\beta,\gamma$ of matrix element $\mathcal{M}_{\alpha\beta\gamma}$ correspond to the choice of polarization tensors of the particles with momenta $p_1,p_2,k$, respectively. Summing over all polarizations gives the total amplitude presented in the equation \eqref{eq:differential decay rate 1}, which is defined as
   \begin{align}
       \left|\mathcal{M}(k,q)\right|^2=\frac{1}{2}\sum_{i,j,k}\left|\mathcal{M}_{ijk}\right|^2.
   \end{align}
   Here, the factor $\frac{1}{2}$ comes from the fact that the two outgoing photons are identical particles.

\bibliographystyle{utphys}
\bibliography{References}


\end{document}